\documentclass{IEEEcsmag}

\usepackage[colorlinks,urlcolor=blue,linkcolor=blue,citecolor=blue]{hyperref}
\usepackage{hyphenat}
\usepackage{draftwatermark}
\SetWatermarkLightness{ 0.9 }
\SetWatermarkText{DRAFT}
\SetWatermarkScale{ 1 }

\jvol{XX}
\jnum{XX}
\paper{8}
\jmonth{April/May}
\jname{Computer}
\pubyear{2020}

\begin{document}


\title{Blockchain Architecture for Auditing Automation and Trust Building in Public Markets}

\author{Sean Cao}
\affil{Georgia State University}

\author{{Lin William Cong}}
\affil{Cornell University}

\author{Meng Han}
\affil{Kennesaw State University}

\author{Qixuan Hou}
\affil{Georgia Institute of Technology}

\author{Baozhong Yang}
\affil{Georgia State University}
\hyphenation{financial}
\hyphenation{account}
\hyphenation{vouchers}
\hyphenation{Verification}
\hyphenation{follow}
\hyphenation{applicable}
\begin{abstract}
Business transactions by public firms are required to be reported, verified, and audited periodically, which is traditionally a labor-intensive and time-consuming process. 
To streamline this procedure, we design FutureAB (\underline{Future} \underline{A}uditing \underline{B}lockchain) which aims to automate the reporting and auditing process, thereby allowing auditors to focus on discretionary accounts to better detect and prevent fraud. 
We demonstrate how distributed-ledger technologies build investor trust and disrupt the auditing industry. 
Our multi-functional design indicates that auditing firms can automate transaction verification without the need for a trusted third party by collaborating and sharing their information while preserving data privacy (commitment scheme) and security (immutability). 
We also explore how smart contracts and wallets facilitate the computerization and implementation of our system on Ethereum. 
Finally, performance evaluation reveals the efficacy and scalability of FutureAB in terms of both encryption (0.012 seconds per transaction) and verification (0.001 seconds per transaction). 
\end{abstract}

\maketitle

\section{Introduction} 

\chapterinitial{F}inancial auditing is a systematic and independent process of examining an organization's financial data, including books, accounts, statutory records, documents, and vouchers to determine if they are accurate and compliant with laws and regulations. 
Verification of counter-party transactions is an essential part of auditing. 
Public firms tend to be large, with a total global market capitalization of \$68.7 trillion. 
Auditing firms handle large quantities of mechanical transaction verification and have limited resources for more sophisticated tasks that require discretionary judgment and expertise. 
Due to the high cost of verification, auditors usually randomize audit samples.
Consequently, traditional auditing is necessarily partial with a considerable potential for misreporting, which in turn erodes investors' trust in public markets \cite{cao2018auditing}. 

Moreover, auditing firms may possess mutually useful information, yet prefer to work independently because (i) clients are reluctant to authorize the sharing of data, which makes it illegal for third parties to do so, especially after regulations such as the General Data Protection Regulation (GDPR); and (ii) traditional infrastructure does not have a mechanism to share data in a cost-efficient way. 
In practice, when verifying transactions, auditors contact the transaction counter-parties either manually or through a third party, which may not always be reliable \cite{cao2018auditing}. 
Collaboration between auditing firms is challenging, primarily due to the lack of a system that is not only secure from hacking, but also scalable and efficient in handling a large user base and multitudinous transactions.

As a potential solution to these problems, in this paper we present FutureAB, a blockchain-based platform for collaborative auditing with advanced privacy protections. 
Thanks to the decentralized nature of blockchain, FutureAB can automate transaction auditing between firms without the need for a trusted third party. 
To ensure the privacy of proprietary data, we have adapted the Pedersen commitment to produce a modified data exchange scheme for detailed transfer of information along with the transactions. 
FutureAB also employs a smart wallet system and smart contracts to further improve efficiency. 
We strengthen the protocol with ledgers to keep track of records with immutability and ensure informational security. 
Information stored on various ledgers makes it simple and easy to detect manipulation attempts.
Finally, we implement our FutureAB system on Ethereum to evaluate its performance. We find that FutureAB is scalable and efficient, with an encryption speed of 0.012 seconds per transaction and verification at 0.001 seconds per transaction.

Our system answers the auditing industry's call for blockchain-based innovation. 
Although all of the Big 4 auditing firms are aware of the importance of blockchain and are devoting vast resources to its development by establishing research labs or providing blockchain services (e.g.,  \cite{Bajpai2017}, \cite{Vetter2018}), it is still unclear how exactly this emerging technology will affect the auditing industry and indeed the auditors themselves. 
While accounting firms' recent efforts center on building in-house blockchain capabilities and services (e.g. \cite{Bajpai2017}, \cite{CNN2018}), our paper demonstrates the possibility of connecting isolated auditing processes while preserving data privacy with blockchain technology. 

The rest of the paper is organized as follows. We review related literature, present the overall design of FutureAB, and provide technical and implementation details. We then perform comprehensive evaluations of the implemented system, before concluding with a discussion of blockchain applications in finance and accounting.

\section{Related Work}\label{sec:2}

\subsection{Collaborative \& Continuous Auditing}
In auditing, collaboration is often identified as a way to reduce costs and improve efficiency.
There are several existing applications designed for collaborative auditing. 
Wu et al. propose an agent-based architecture to increase the frequency of periodic audits \cite{wu2008agent}. 
This scheme emphasizes efficiency in continuous auditing, but does not address privacy concerns. 
Sachar et al. present a framework based on the concept of an ``audit warehouse" that enables central, tool-supported auditing of cross-enterprise business processes \cite{paulus2007collaborative}. 
Chen et al. develop a collaborative continuous-auditing model relying on XML and Web Service technologies under service-oriented architecture environments \cite{chen2007collaborative}. 
A complex protection profile is required to ensure data security in these two frameworks. Wang et al. propose a secure cloud storage system to support secure public auditing and introduce a third party to check the integrity of data \cite{wang2010privacy}. 
However, the integrity and reliability of the third party is not guaranteed. 
Our blockchain architecture circumvents the aforementioned issues by implementing a decentralized verification mechanism. 

\subsection{Smart Contract}
Nick Szabo proposes ``smart contracts," computer protocols that can automatically execute the terms of a contract, facilitate and verify the performance of a contract, interact with other contracts, make decisions, store data, or send data to others \cite{szabo1997formalizing}. Many smart contract platforms are now emerging, including Ethereum, Hyperledger, and Corda. 
The FutureAB platform takes full advantage of smart contracts to further minimize human error and improve efficiency in auditing. 

\subsection{Commitment Schemes}
The Pedersen commitment is a  commitment scheme based on cryptographic hash functions 
\cite{pedersen1991non}. 
A commitment scheme allows the sender to commit to a choice while hiding their selections from other receivers. Commitment schemes are widely used in blockchain applications to preserve privacy. There are several examples of this tactic in recent literature. For example, Knirsch et al. propose using commitment schemes in electronic vehicle charging \cite{knirsch2018privacy}; Zhang et al. propose BCPay, a blockchain-based fair payment framework for outsourcing services in cloud computing \cite{zhang2018blockchain}; 
Xu et al. discuss the potential of commitment schemes for enabling sharing economies \cite{xu2017enabling}.) 

FutureAB leverages the Pedersen commitment with specific adjustments to guarantee information security. Specifically, our proposed application would ensure the suppression of auxiliary information not directly related to transactions in order to protect participants' data privacy.
For example, if company $A$ in Atlanta shipped 1,000 units to company $B$ in Phoenix, which arrived on 1/1/2020, the transaction record would contain affiliated information that might be potentially useful to competitors, such as date and type of product. Our goal is to provide not only a platform for the auditing process but also a mechanism that would prevent the disclosure of unnecessary information, which would provide an incentive for inter-company auditor collaboration. 

\section{System Design} \label{sec:3}

\subsection{Pain Points of Current Auditing Processes}
Traditionally, the auditing process of each company is independent. Several issues arise:

\textbf{High cost and low efficiency:} Auditors of one company have to request transaction records from counter-parties and manually verify the information, which is a labor-intensive process.

\textbf{Failure to fully utilize all information:} Reducing auditing sample size is a common way to reduce costs. 
However, Cao et al. underscore that the sample size correlates with the quality of auditing \cite{cao2018auditing}, the failure of full information utilization therefore negatively affects the end result.

\textbf{Fraudulent reports:} 
Failure to use all information in auditing also creates a greater potential for fraud or misreporting. Companies may overstate earnings to boost their stock market valuation.

\textbf{Privacy and access:} 
A platform for auditors to share transaction information might reduce the cost and improve the efficiency and quality of the auditing process. 
However, companies are reluctant to reveal proprietary information to others, especially their competitors.

\subsection{Business Process Design}
\begin{figure}
	\centerline{\includegraphics[width=18.5pc]{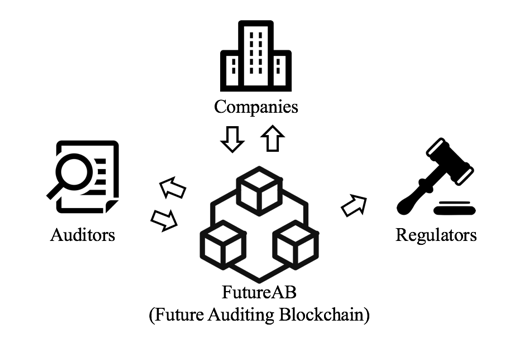}}
	\caption{Different roles in FutureAB.}
	\label{sys1}
\end{figure}

FutureAB addresses the aforementioned challenges. 
The platform focuses on auditing transaction-based accounts and, as shown in Figure \ref{sys1}, assists auditors in investigating mismatched transactions, companies being audited, and regulators who oversee these processes.
The whole system is permission-based, meaning that permission could be granted by the committee of the participants, such as the auditing association or the Company Public Accounting Oversight Board (PCAOB).
The public key or address used within the blockchain would still be limited to the members of the system and would not be ``public.'' All historical transactions are stored locally in the auditor members' proprietary databases.

\begin{figure}
	\centerline{\includegraphics[width=18.5pc]{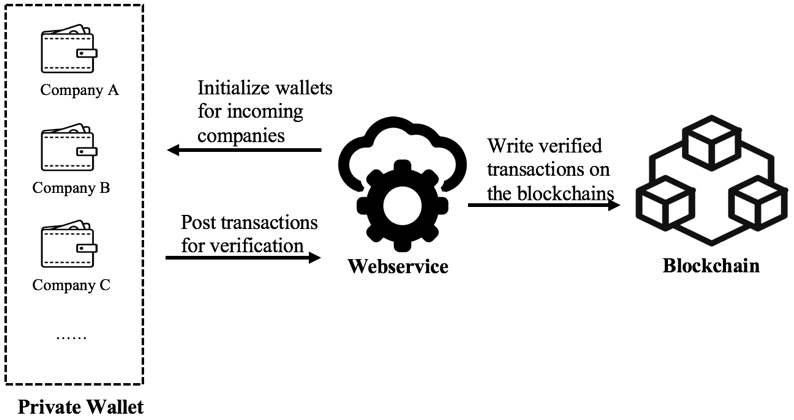}}
	\caption{Business process design of FutureAB}
	\label{sys2}
\end{figure}

 Figure \ref{sys2} displays the three major components of FutureAB. 

\textbf{Private wallet:} 
Under the proposed architecture, each company possesses a private electronic wallet, called ABWallet, which stores its public addresses, private keys, and confidential information for encryption within the system. We next discuss how the wallet is generated, how addresses and keys are managed, and how the information is stored.

\textbf{Web-service:} Auditors and regulators can access a public web-based application to perform tasks such as reviewing mismatched transactions.
Smart contracts are deployed here to pair posted transactions for verification and then write verified transactions onto the blockchain. 

\textbf{Blockchain:} Any key holder can use their private key to sign the verified transactions. The resulting signature is then recorded on the blockchain for peers to verify. 

The business process of FutureAB is as follows.

\begin{figure}
	\centerline{\includegraphics[width=18.5pc]{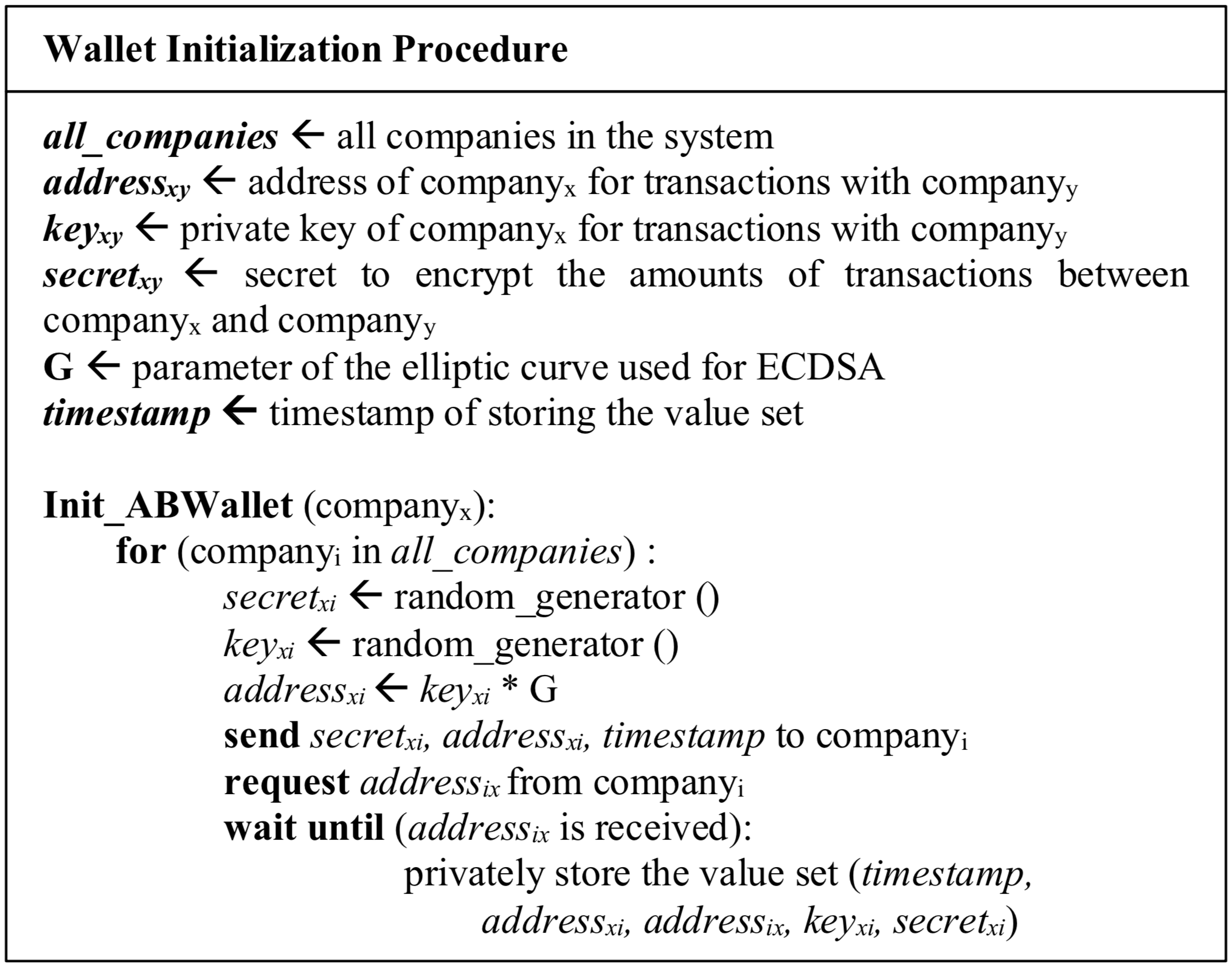}}
	\caption{Wallet initialization procedure.}
	\label{p1}
\end{figure}

1) \textit{Initialize ABWallet for incoming companies.} A company can join the system by requesting access via the website. 
Once the access is granted, the incoming company can download ABWallet, which generates, stores, and manages public addresses, private keys, and commitment secrets for further activities on FutureAB. 
The organizations overseeing the auditing should be the ones who monitor the blockchain system and respond to the companies' access requests. As mentioned earlier, the PCAOB could be a good candidate to maintain the auditing blockchain. An alternative would be an alliance of major auditing firms.

The wallet initialization procedure is presented in Figure \ref{p1}. When $company_x$ joins the system, the company selects a set of other companies they often work with. In response, ABWallet generates distinct public addresses, private keys, and commitment secrets for the selected companies, sends public addresses and commitment secrets to the corresponding companies, and then requests addresses from the counter-party, $company_y$. 
FutureAB provides a company gallery, which sorts companies into different categories, such as a set of media companies, a set of healthcare companies, and a set of retail companies. The incoming company can pick several sets as its potential counter-parties. 
This feature enables the wallet initialization process to be more efficiently achieved in small batches. When posting transactions, ABWallet can automatically generate addresses and request counter-parties’ addresses if one of the counter-parties does not yet exist in the wallet. 

Once the addresses are received, value sets (timestamp, $address_{xy}$, $address_{yx}$, $secret_{xy}$, $key_{xy}$) are stored privately in the wallet of $company_x$. At the same time, $company_y$ also privately stores the value sets (timestamp, $address_{yx}$, $address_{xy}$, $secret_{xy}$, $key_{yx}$). We use the Elliptic Curve Digital Signature Algorithm (ECDSA) as our signature scheme.

\begin{figure}
	\centerline{\includegraphics[width=18.5pc]{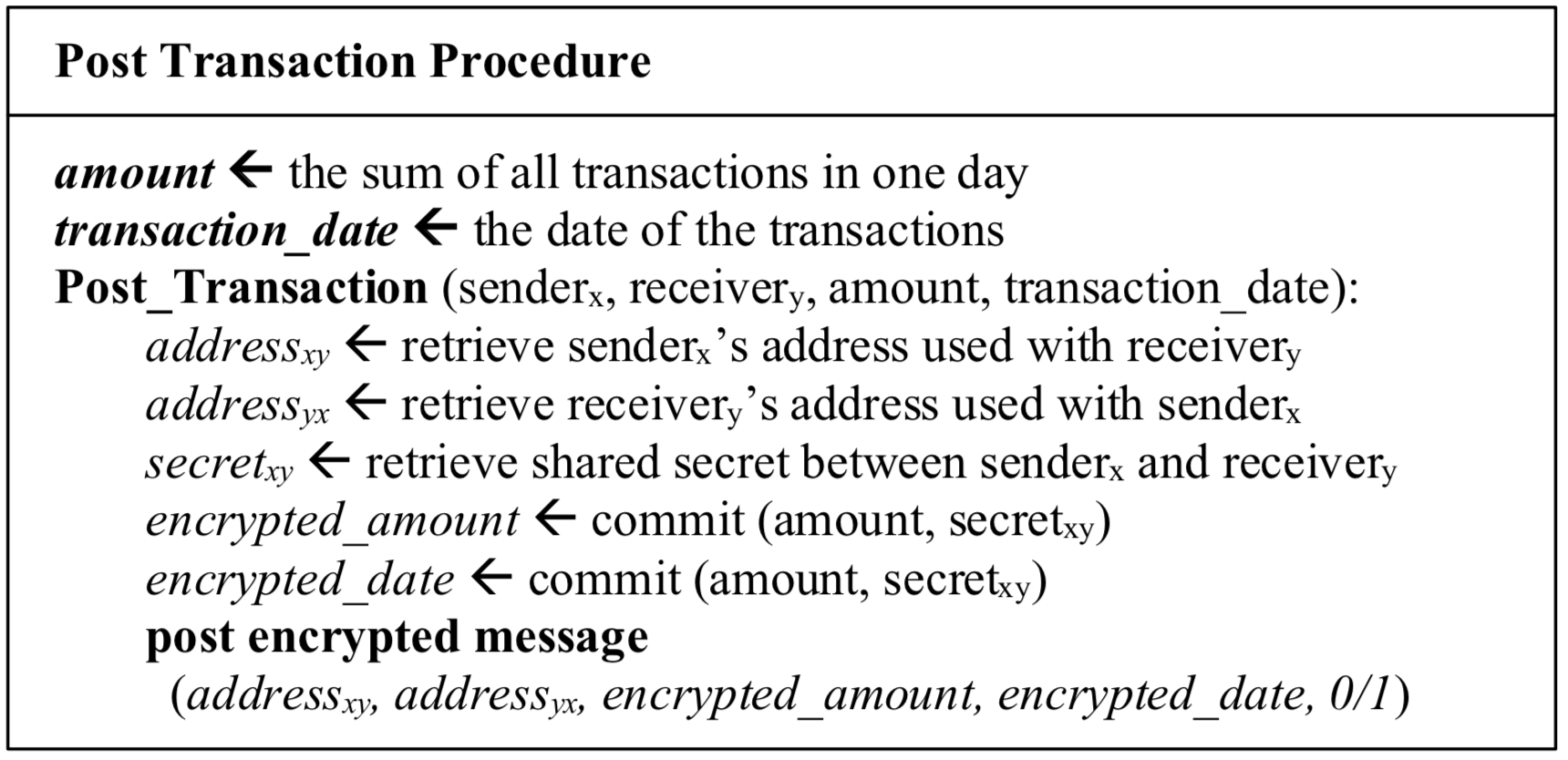}}
	\caption{Post transaction procedure.}
	\label{p2}
\end{figure}

2) \textit{Post transactions for verification.} 
ABWallet generates signatures and associated signed messages so that posting transactions on the web service will not compromise digital security. The post transaction procedure is presented in Figure \ref{p2}. The messages are structured as the sender address, the receiver address, the amount, the date, and 0 or 1. `0' indicates the posted transaction is from the sender; otherwise, the last digit of the message is `1.' Once the message is posted successfully, the status of the message is labeled as ``pending,'' which means that it is pending verification. 

Beyond basic information, companies are encouraged to post details of a transaction and hide the information from the public with a Pedersen commitment. If a discrepancy is spotted and auditors are notified, the auditors can request the commitments to be opened in order to review the details of the transactions. The auditors can also contact the corresponding companies if more information is needed for investigation.

3) \textit{Verify posted transactions.} Both counter-parties should post the transaction and encrypted messages with the same sender address, receiver address, and commitment secret. If there are multiple transactions between two companies within the same day (based on GMT), the FutureAB will take the sum of these so that there is only one transaction between two companies per day. The web service consistently attempts to pair up two messages. A pair is defined as two messages with the same sender address, receiver address, and date. As described above, the last digit should be ``0'' or ``1.''

\begin{figure}
	\centerline{\includegraphics[width=16.5pc]{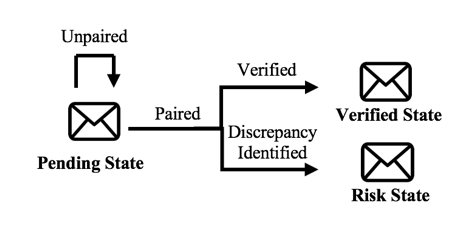}}
	\caption{Three possible states of a posted message}
	\label{sys3}
\end{figure}

There are three possible states of one posted message as shown in Figure \ref{sys3}. \textbf{Verified:} If the message is paired with another and two messages are identical except for their respective last digits, the transaction in the message is verified by both involved parties and can be written on the blockchain as a permanent record. \textbf{Risk:} 
A discrepancy is identified when a pair of messages contain different amounts. We label the pair as a ``risk'' to notify auditors to trigger an investigation. 
Being freed from mechanical transaction verification, auditors can focus on discretionary accounts where knowledge is indispensable. \textbf{Pending:} If only one involved party posts the transactions, this asymmetry may produce messages that cannot be paired. 

\section{\bfseries{Technical Details}} \label{sec:4}

\subsection{ABWallet}
We introduced ABWallet to allow each company to generate and manage value sets. ABWallet is also responsible for communicating the latest public addresses and commitment secrets between companies in order to ensure the synchronization of information. Whenever a company initiates the process of posting transactions, ABWallet retrieves the latest value sets and encrypts the transactions.

A member joins the system by downloading ABWallet and starting the wallet initialization procedure discussed in the previous section. Access to ABWallet should be kept private and secure. 
The only information flowing among different members' wallets should be public addresses and commitment secrets of counter-parties, and the only information flowing from a wallet to the web service should be signed messages. 

For FutureAB, we propose using different addresses for transactions with different companies in order to preserve anonymity. 
We also recommend frequently generating new addresses in order to hide companies' identities in the posted transactions.  

\begin{figure*}
	\centerline{\includegraphics[width=33.5pc]{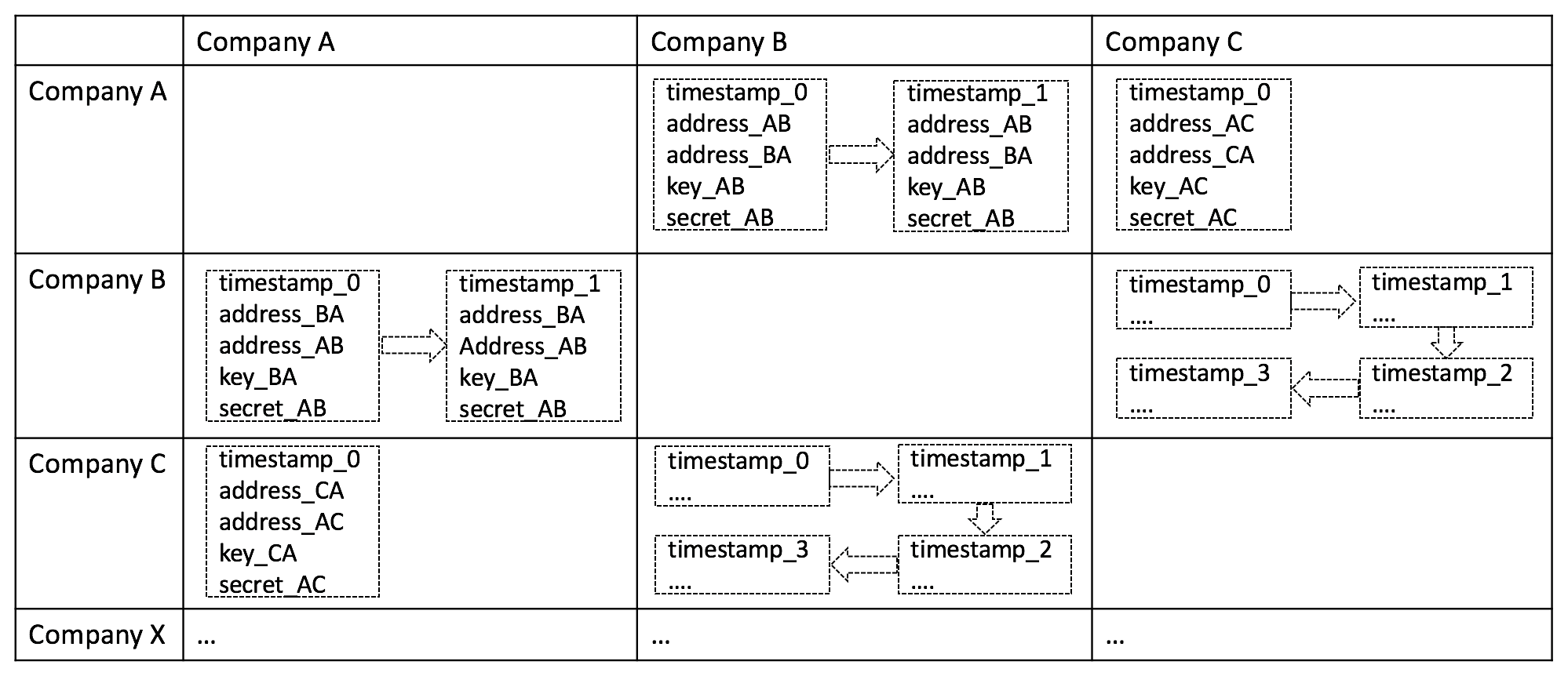}}
	\caption{Information Stored in ABWallet}
	\label{t1}
\end{figure*}

As shown in Figure \ref{t1}, each wallet of the corresponding company is a row. For instance, Company A only has access to the first row in the table. When Company B generates a new value set for transactions with Company A and updates the first cell in the second row, Company A will be notified with new $address_{BA}$ and $secret_{BA}$ and a new value set will be appended to the second cell in the first row. ABWallet helps companies generate value sets and maintain up-to-date information so that verification procedures can be executed quickly and correctly.


\subsection{Smart Contracts}
Once smart contracts are compiled and migrated, the web service can implement the smart contracts when certain conditions are satisfied. In our setup, the smart contract is triggered when a new message is posted. It is executed to pair messages and then write verified messages on the blockchain.

Compared to a traditional auditing system, smart contracts considerably reduce manual effort and costs in verifying transactions because they are code-based and run live on the Internet at a low cost.

\subsection{Commitment Schemes}
In FutureAB, only two involved parties share the secret 
%
%
to decrypt the message, meaning that the message is hidden from all other parties on the blockchain. 
In the meantime, both participating parties use the same secret to execute a transaction which is then posted on the web service, indicating that both parties can no longer change what is committed. Note that FutureAB can accommodate transactions involving multiple parties.


This effort helps preserve the integrity of the content without disclosure; when details are inquired, the commitment could guarantee the trustworthiness of the committed content. 
We adapted the Pedersen commitment scheme in the design of FutureAB. 
The Pedersen commitments have the hiding property, which indicates that the commitment reveals nothing about the message. 
Additionally, Pedersen commitments are homomorphic, which facilitates the quick generation and verification of transactions on FutureAB by making it possible to combine commitments. If $cm_1$ and $cm_2$ are two commitments to values $v_1$ and $v_2$, using commitment randomness $r_1$ and $r_2$, then $cm:= cm_1 * cm_2$ is a commitment to $v_1 + v_2$ using randomness $r_1 + r_2$. 
The commitment could preserve the security of certain information related to the transaction——such as transaction descriptions, quantities of products, and exchanged strategies, etc.——to the highest extent, which would encourage a larger number of participates to join and collaborate on the system.

\section{Implementation Details} \label{sec:5}
\subsection{Development Environment}
The main components of FutureAB are ABWallet, the web application, and the blockchain. The Ethereum blockchain is a public blockchain that allows users to perform Turing-complete calculations (smart contracts). 
The Ethereum protocol has an average block time of 15s and charges small transaction fees for the processing of smart contracts. 
Ethereum satisfies all the current requirements of FutureAB. The confirmation time and other specific requirements of FutureAB should be the subject of future research. 

The web application interacts with smart contracts on the Ethereum blockchain. 
The web app is written in JavaScript and HTML5, using the Truffle development framework. The framework enables JavaScript bindings for the smart contract and includes libraries such as web3.js that facilitate communication between the web app and the Ethereum client. 

\subsection{Rewards Program} 
FutureAB is designed to allow an optional rewards programs implemented to attract more companies and auditors. 
A rewards program can be introduced to motivate every party in the system to actively and continuously post their transactions in order to achieve collaborative and continuous auditing as an expected outcome of FutureAB. 
The rewards program can also motivate companies to mine and sign the verified transactions so that the records can be permanently written on the blockchain.

\section{Evaluation} \label{sec:6}
To the best of our knowledge, FutureAB is one of the first platforms that can support collaborative and continuous auditing on the blockchain without compromising data privacy.
%
%
%

\begin{figure}
	\centerline{\includegraphics[width=16.5pc]{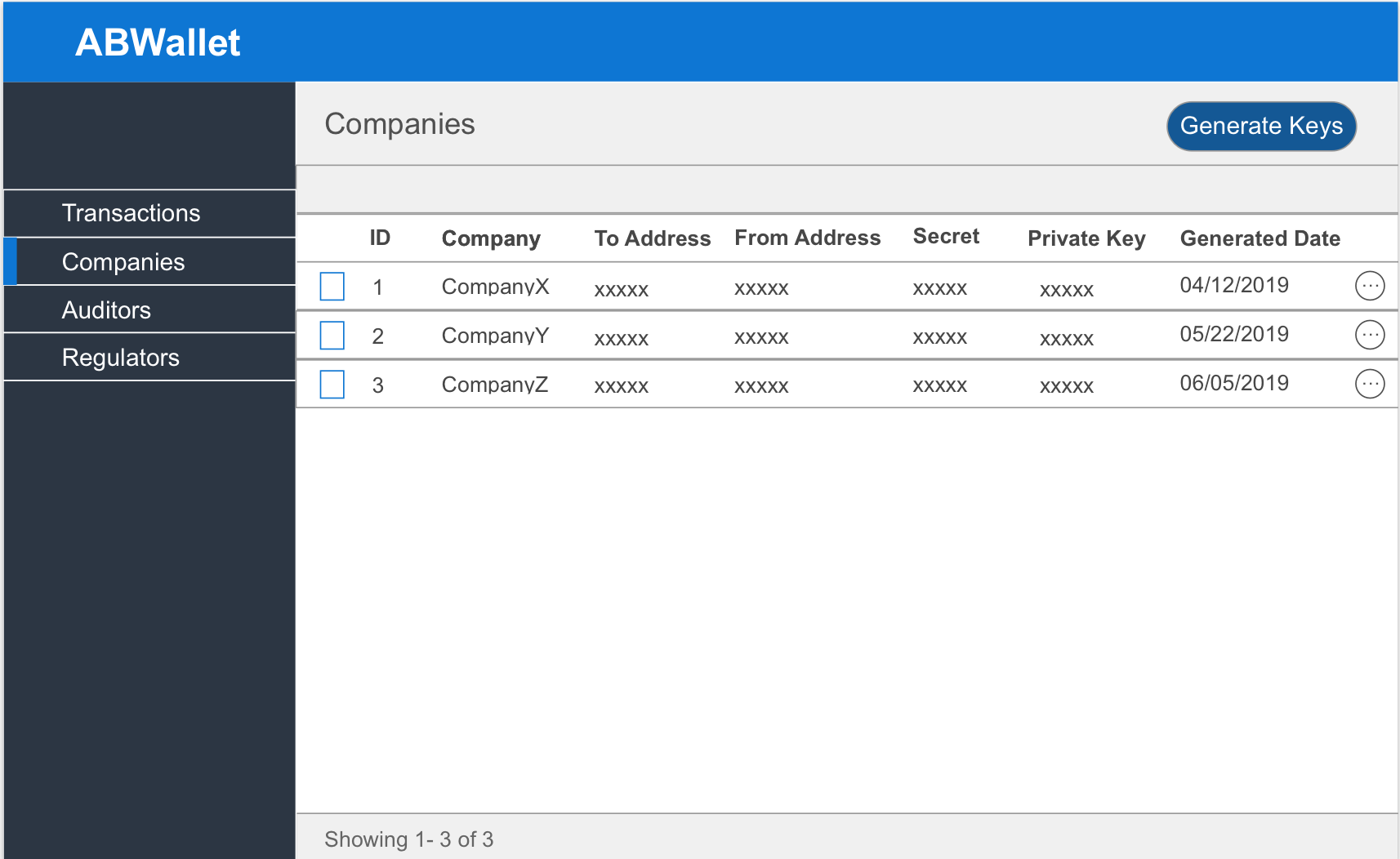}}
	\centerline{\includegraphics[width=16.5pc]{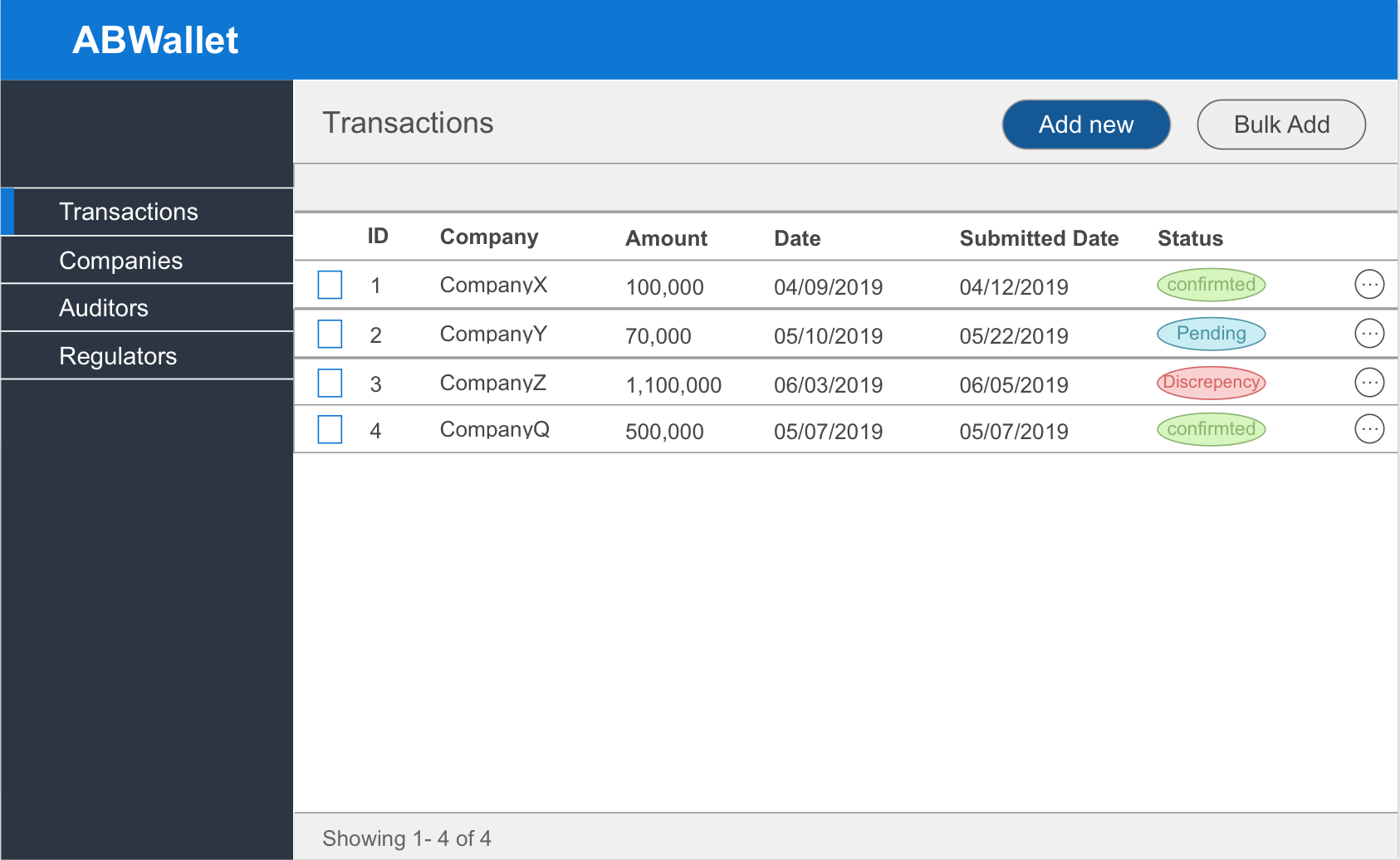}}
	\caption{User interface design of ABWallet.}
	\label{fig5}
\end{figure}

Once ABWallet is downloaded and initialized for ${company_x}$, users from ${company_x}$ can review and manage partner companies’ addresses, keys, and secrets via the list view in Figure \ref{fig5}. 
The context menu on each row allows users to generate new sets of values, to request new addresses from their counter-parties, or to view the transaction history. Users can also view transaction details in the list view.
The ``Bulk add'' button on the top right of the screen allows users to upload multiple transactions with an Excel sheet template. 
The status of each transaction is indicated in the status column. 

\begin{figure}
	\centerline{\includegraphics[width=15.5pc]{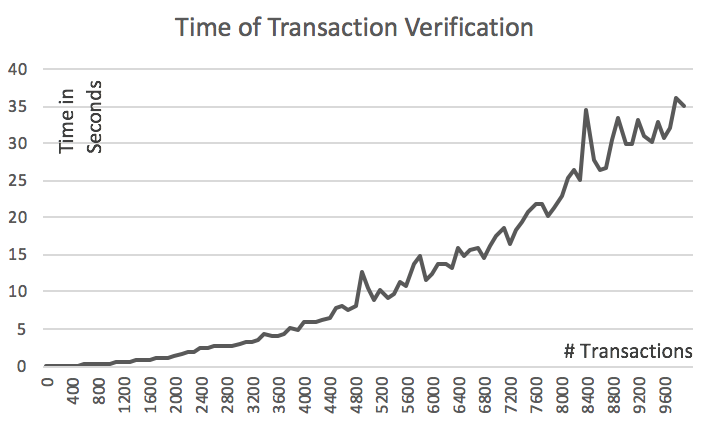}}
	\caption{Time of transaction verification.}
	\label{fig9}
\end{figure}

We ran the simulation of wallet initialization and the transaction encryption on a 2.2 GHz Intel Core i7 machine with 16GB of RAM and 1600 MHz CPUs. This hardware is only used for simulation and performance evaluation. Since FutureAB is a distributed blockchain, the communication traffic and the resource usage for each node are much lower than those on a single machine simulation once deployed.
It takes 0.096 seconds on average to set up one value set for one counter-party. It takes about 16 minutes to set up the wallet for a new company when there are 10,000 selected counter-parties. 
It takes 0.021 seconds on average to encrypt one transaction. Less than 7 minutes are needed to encrypt 20,000 transactions. 
As shown in Figure \ref{fig9}, the current system takes less than one minute to verify 10,000 transactions. 
As such, we believe that FutureAB can support real-time posting, i.e., companies encrypting and posting transactions simultaneously with, or a short period of time after, the occurrence of the events.

\section{Discussion and Conclusion} \label{sec:6}

In this paper, we presented a blockchain architecture to automate collaborative and continuous auditing and, in so doing, to build trust in public markets.

Blockchain is one of the most influential emerging technologies in the past decade. The distributed nature of blockchains achieves peer-to-peer communication and allows auditing collaboration and financial reporting without relying on a trusted third party (\textbf{Decentralization}). Blockchains naturally provide immutability, which guarantees that once an accounting activity is recorded, nobody, including the owner of the business, can arbitrarily change the records (\textbf{Immutability}). Moreover, smart contracts use protocols and algorithms to digitally and automatically facilitate, verify, or perform a contract between two parties within a blockchain  (\textbf{Automation}). The use of encryption techniques protects proprietary information while ensuring certain messages can be recorded on a public blockchain without compromising privacy (\textbf{Encryption}). 

Although blockchains with encryption boast many exciting features, there are still several limitations. The transaction verification needs to pass a certain level of synchronization on the whole peer-to-peer system, which may result in a delay of seconds to minutes, not to mention the extra storage required. 
Fortunately, the rapid developments of hardware, consensus algorithms, and storage technology mitigate these concerns, especially for financial reporting and auditing processes that do not require a millisecond-level performance. 

More generally, our proposed architecture provides an alternative way to achieve privacy-preserved information exchange. Besides auditing, the system design could be applied to many other fields for information collaboration, including banking, insurance, and even healthcare.

\section*{Acknowledgement}
We would like to thank Jasmine Cheng and Fahad Saleh for insightful feedback and comments.

\begin{IEEEbiography}{Dr. Sean Cao} is an Assistant Professor at the School of Accountancy, Robinson School of Business, Georgia State University, Atlanta, GA, USA. He received his Ph.D. in Accountancy from the University of Illinois at Urbana-Champaign. Dr. Cao’s research interests include FinTech applications in capital markets. His research has been published in leading journals such as the Accounting Review, Journal of Accounting Research, Contemporary Accounting Research, and Accounting Horizon. He teaches Ph.D. seminars in financial accounting and undergraduate courses in both financial accounting and taxation.
\end{IEEEbiography}

\begin{IEEEbiography}{Dr. Will Cong} is the Rudd Family Professor of Management and associate professor of Finance. He is an associate editor at Management Science and directs the FinTech Initiative at Cornell University. He received his Ph.D. in Finance and M.S. in Statistics from Stanford University. His research interests include Economic Data Science, FinTech, and Information Economics. He has published in various leading journals and has been invited to hundreds of research institutions, including the IMF and Federal Reserve Banks.
\end{IEEEbiography}

\begin{IEEEbiography}{Dr. Meng Han} is an Assistant Professor in the College of Computing and Software Engineering at Kennesaw State University. He received his Ph.D. in Computer Science from Georgia State University. He is currently an IEEE member and an IEEE COMSOC member. His research interests include data-driven intelligence, FinTech, and Blockchain technologies.
\end{IEEEbiography}

\begin{IEEEbiography}{Ms. Qixuan Hou} is a Master student in Analytics, received her B.S. in Computer Science and Mathematics from Georgia Institute of Technology. Her research interests are in the area of data science, with experience executing data-driven solutions. She is interested in building innovative, intelligent systems to deliver insights and implement action-oriented solutions for complex real-life problems. 
\end{IEEEbiography}

\begin{IEEEbiography}{Dr. Baozhong Yang} received his Ph.D. from Stanford University and MIT. He organized the inaugural and second GSU-RFS FinTech Conferences and has served on the Program Committee of many conferences. His research interests include FinTech and Investments. He has published in leading journals such as the Journal of Finance, Journal of Financial Economics, Review of Financial Studies, and Management Science.
\end{IEEEbiography}

\end{document}